\documentclass[journal]{IEEEtran}
\usepackage{amsmath,amsfonts}
\usepackage{algorithmic}
\usepackage{algorithm}
\usepackage{amsmath}
\usepackage{array}
\usepackage[caption=false,font=normalsize,labelfont=sf,textfont=sf]{subfig}
\usepackage{textcomp}
\usepackage{stfloats}
\usepackage{url}
\usepackage{verbatim}
\usepackage{cite}
\usepackage{multirow}
\usepackage{bigstrut}
\usepackage{amsfonts}
\usepackage[section]{placeins}
\usepackage{bbm}

\usepackage{algorithmic}
\usepackage{algorithm}

\usepackage{graphicx}
\ifCLASSINFOpdf
\else
\fi
\makeatletter
\newenvironment{breakablealgorithm}
  {
   \begin{center}
     \refstepcounter{algorithm}
     \hrule height.8pt depth0pt \kern2pt
     \renewcommand{\caption}[2][\relax]{
       {\raggedright\textbf{\ALG@name~\thealgorithm} ##2\par}%
       \ifx\relax##1\relax 
         \addcontentsline{loa}{algorithm}{\protect\numberline{\thealgorithm}##2}%
       \else 
         \addcontentsline{loa}{algorithm}{\protect\numberline{\thealgorithm}##1}%
       \fi
       \kern2pt\hrule\kern2pt
     }
  }{
     \kern2pt\hrule\relax
   \end{center}
  }
\makeatother

\begin{document}

\title{Self-Updating Vehicle Monitoring Framework Employing Distributed Acoustic Sensing towards Real-World Settings}
%
%
%

\author{Xi~Wang,
        Xin~Liu,
        Songming~Zhu,
        Zhanwen~Li
        and Lina~Gao
\thanks{The authors are with Department of Earth Sciences, University of Hong Kong. Corresponding author: Xin Liu.}
}

%
%

\markboth{Arxiv}%
{}
%



\maketitle

\begin{abstract}
The recent emergence of Distributed Acoustic Sensing (DAS) technology has facilitated the effective capture of traffic-induced seismic data.
The traffic-induced seismic wave is a prominent contributor to urban vibrations and contain crucial information to advance urban exploration and governance. 
However, identifying vehicular movements within massive noisy data poses a significant challenge.
In this study, we introduce a real-time semi-supervised vehicle monitoring framework tailored to urban settings.
It requires only a small fraction of manual labels for initial training and exploits unlabeled data for model improvement.
Additionally, the framework can autonomously adapt to newly collected unlabeled data. 
Before DAS data undergo object detection as two-dimensional images to preserve spatial information, we leveraged comprehensive one-dimensional signal preprocessing to mitigate noise.
Furthermore, we propose a novel prior loss that incorporates the shapes of vehicular traces to track a single vehicle with varying speeds. 
To evaluate our model, we conducted experiments with seismic data from the Stanford 2 DAS Array. 
The results showed that our model outperformed the baseline model Efficient Teacher and its supervised counterpart, YOLO (You Only Look Once), in both accuracy and robustness.
With only 35 labeled images, our model surpassed YOLO's mAP 0.5:0.95 criterion by 18\% and showed a 7\% increase over Efficient Teacher. 
We conducted comparative experiments with multiple update strategies for self-updating and identified an optimal approach. 
This approach surpasses the performance of non-overfitting training conducted with all data in a single pass.

\end{abstract}


%
\IEEEpeerreviewmaketitle

\section{Introduction}

%
%
%
%

\IEEEPARstart{B}{y} shifting focus from traditional natural seismic events to human-induced vibrations, urban seismology helps with urban management and rises in prominence within applied geoscience \cite{diaz2017urban,maciel2021urban}.
In urban environments, traffic-induced seismic vibrations provide rich data with implications to enable continuous monitoring of traffic and infrastructure \cite{zhang2019near}.
However, it is still quite challenging for the dense deployment of traditional seismometers due to cost and spatial constraints \cite{spica2020urban,park2020machine}.
In addressing these concerns, the economical, anonymous, and flexible geophysical sensor system, Distributed Acoustic Sensing (DAS), presents a promising alternative \cite{liu2019vehicle,wang2020rose}.

DAS is an emerging geophysical sensing system that repurposes telecommunication-grade cables into sensitive sensors \cite{parker2014distributed,zhan2020distributed,cheng2024photonic,li2018pushing}.
Comprised of highly maintainable and flexible optical fibers, DAS offers high-resolution seismic measurements at a low cost over tens of kilometer scales \cite{dumont2020deep,hileman2021development}.
It even reaches sites previously inaccessible (e.g., volcanoes, glaciers, and urban areas) \cite{lindsey2021fiber}.
In seismology, DAS is adapted to perform both active source imaging  \cite{mateeva20174d,daley2013field,mateeva2014distributed} and ambient noise imaging \cite{lindsey2019illuminating,song2021sensing,chen2024surface} for its long duration and high resolution, showing a distinct advantage over traditional seismometers.
Analogous to seismological applications, object monitoring with DAS involves detecting seismic events caused by significant ground vibrations. 
These vibrations are widespread regardless of weather conditions, and the monitoring process is entirely anonymous without privacy concerns.

Although DAS is highly effective for capturing vehicle vibration from the sensor attributes perspective, tracking accurate vehicle traces is still a considerable challenge.
The primary challenge is various types of strong noise. 
DAS is highly sensitive to minor vibrations, and environmental vibration can introduce substantial noise, overwhelming the signals to be extracted \cite{yuan2023spatial}.
In addition, the vast quantities of data and the diverse shapes of targets pose significant detection complexities.

Many studies of DAS applications in other related areas have laid the groundwork for the vehicle detection task.
To suppress noise in DAS data, Chen \emph{et al.} proposed an integrated denoising framework for geothermal data, earthquakes, and microseismic events, combining multiple methods for different types of noise \cite{chen2023denoising}.
Machine learning is also widely used in denoising due to its powerful generalization capabilities \cite{zhao2020distributed,zhao2022coupled,yang2023denoising,cheng2023simultaneous}.  
For the challenge of detecting targets with varying shapes, the network YOLO is popular for its real-time and high-precision detection.
It has been applied to microseismic events \cite{stork2020application} and perimeter security air-ground events \cite{wang2024intelligent}.
Besides, convolutional long short-term memory networks are also increasingly utilized for object detection \cite{li2020anti,li2020fiber,rahman2024remote}.
Wang \emph{et al.} proposed a track detection method with semi-supervised learning to address the challenge of large DAS datasets that cannot be fully labeled \cite{wang2022semi}.
For similar purposes, semi-supervised and self-supervised learning have been applied to denoising \cite{zhao2023self, zhao2023sample2sample} and arrival-time picking \cite{zhu2023seismic} to avoid complete labeling of the entire dataset.
However, vehicle detection is more challenging due to the significant influence of vehicle type, speed, road conditions, and the presence of other objects on the road. 
Therefore, more fine-grained and vehicle-specific frameworks need to be developed.

In particular, research for vehicular trace extraction has already begun to address the previously mentioned issues.
As multi-channel one-dimensional (1D) signals, DAS data allow for two processing approaches: analyzing individual time series signals or two-dimensional (2D) images.
Direct 1D signal processing effectively extracts high-frequency amplitude characteristics for monitoring.
Lindsey \emph{et al.} employed the classic seismological algorithm STA/LTA (Short-Time Average over Long-Time Average) for seismic event trigger picking, treating vehicular-induced vibrations as seismic events \cite{lindsey2020city}.
Furthermore, the beamforming method\cite{van2021next} and various energy-based methodologies are also employed to extract vehicular traces \cite{liu2019vehicle,wang2022urban}.
However, disregarding the spatial information of these channels inherently limits the reliability of the monitoring.

Considering the spatial information, 2D image processing achieves more satisfactory results.
The Hough Transform is the most frequently employed method, and it is an algorithmic technique that excels in detecting straight lines \cite{catalano2021automatic,corera2023long,litzenberger2021seamless}.
However, the precondition that vehicular traces are straight lines with constant velocity is unrealistic over extended distances in real-world scenarios.
Building on traditional methodologies, deep learning approaches show better generalization capabilities and accuracy
\cite{yuan2023spatial,van2022deep}.
Ye \emph{et al.} adapted YOLO to discern traces that extend beyond mere linear forms \cite{ye2023traffic}.
However, the manual labeling cost of 10,000 images is time-consuming, and updating the model with newly acquired DAS data remains challenging.

To address the challenges posed by DAS data's huge volume and high noise levels, we designed a comprehensive semi-supervised framework for real-time vehicle monitoring with DAS.
Our framework includes preprocessing and deep learning detection components.
Given the advantages of treating DAS data as one-dimensional signals and two-dimensional images, we combined these characteristics and proposed a preprocessing workflow that captures both spatial and amplitude information features.
Regarding the detection method, we employed a semi-supervised variant of the widely-used supervised model YOLO \cite{redmon2016you,peng2023amflw} and Efficient Teacher \cite{xu2023efficient} for vehicle monitoring.
By embedding the shape prior knowledge of traces into the loss function, we directed the gradient descent during training and effectively addressed the low accuracy issue stemming from sparsely labeled data.
Concurrently, we converted from traditional offline training to an automated online system that approximates real-life settings.
It utilizes only recently collected data and eliminates the need for further labeling.
Detection performance and various update strategies were validated through experiments.

The main contributions of our method can be summarized as follows:
\begin{itemize}
\item[1)]
We applied semi-supervised learning to DAS vehicle monitoring, significantly decreasing the number of required labeled samples.
\item[2)]
We refined the DAS vehicle detection model to facilitate continuous self-enhancement, enabling it to update itself autonomously with newly acquired data.
\item[3)]
We introduced an innovative GPU-friendly shape prior loss that mitigates erroneous detection without increasing the data volume.
\item[4)]
We developed a data preprocessing workflow for DAS trace detection, integrating one- and two-dimensional data features to enhance trace clarity.
\end{itemize}

The rest of this article is organized as follows:
Section~\ref{sec_pre} presents the DAS dataset from the Stanford 2 Array and a detailed description of our data preprocessing workflow. 
Section~\ref{sec_net} describes the architecture of the network employed in our study, highlighting the novel shape prior loss.
Section~\ref{sec_exp} details the comparative experiments conducted to validate the detection performance and the selection of our self-update strategy. 
Section~\ref{sec_dis} discusses the results and future research directions. 
We conclude the final section, Section~\ref{sec_con}.

\begin{figure}[!t]
\centering
\includegraphics[width=3in]{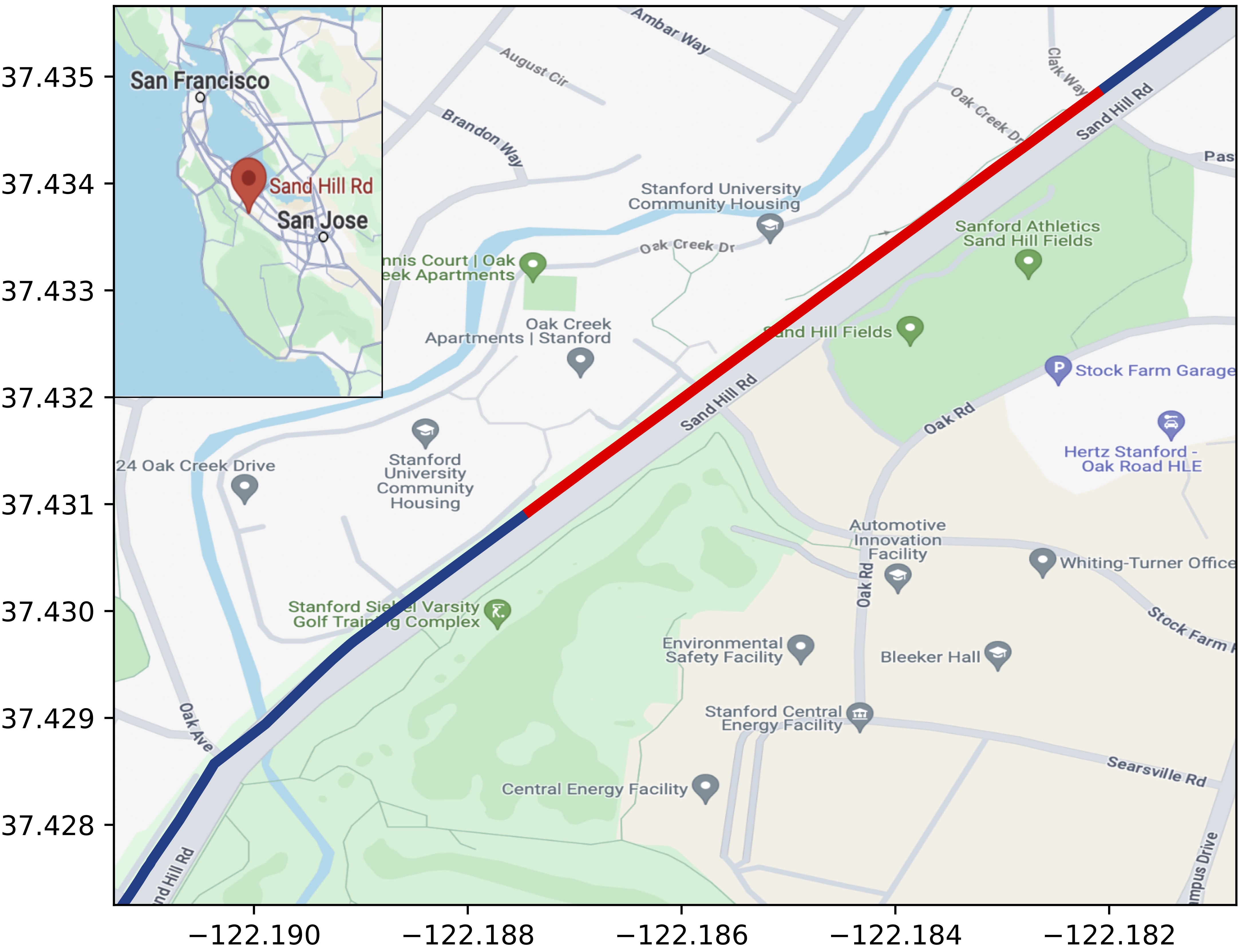}
\caption{Map of DAS Stanford 2 Array. 
In this study, the segments in use are denoted by red lines, whereas the unused ones are marked with blue.}
\label{fig_map}
\end{figure}

\section{Data and Preprocessing}\label{sec_pre}
\subsection{Dataset}
In this study, we utilized DAS recordings of Stanford 2 Array (Sand Hill Road Array) to test the proposed monitoring method \cite{spica2023pubdas}. 
From Stanford Hospital to SLAC National Accelerator Laboratory, the Stanford 2 Array contains raw data sampled initially at 250 Hz between March 1 and 14, 2020.
Data volume is 101 GB per day and 1.4 TB in total.
In contrast to previous experiments conducted on limited-scale internal roads\cite{liu2019vehicle}, the Stanford 2 Array continuously collected data along real-world urban roads for a sufficiently long duration.
Therefore, it is representative for our experiment. 
Due to segment-based monitoring in this workflow, an approximately 600-meter recording near Stanford Hospital lasting 14 days was selected for the experiment, shown in Fig.\ref{fig_map}.

\subsection{Preprocessing Workflow}
\begin{figure*}[!t]
\centering
\includegraphics[width=6.4in]{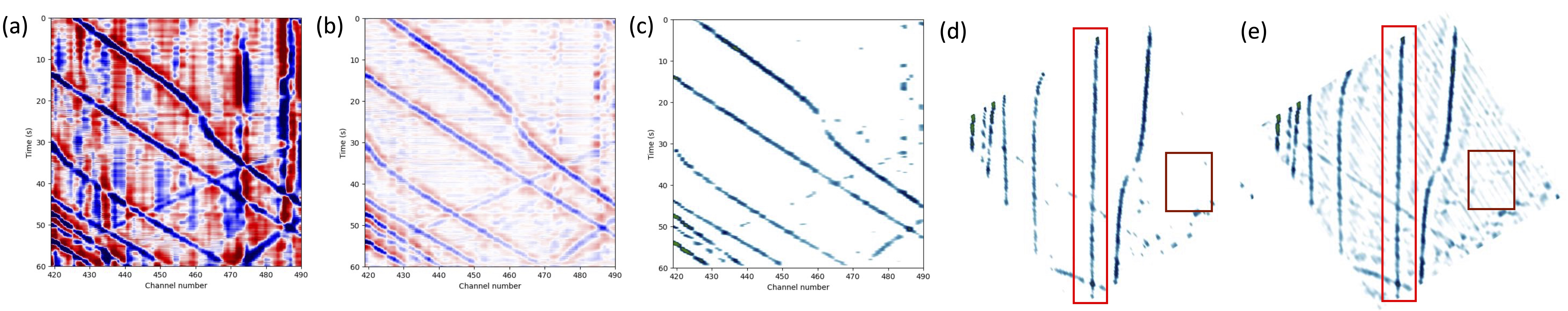}
\caption{Stages of data preprocessing. (a) Raw data with only detrending. The horizontal axis represents the channel number, while the vertical axis corresponds to the time. (b) Images with model-based signal filtering. 
(c) The outputs of STA/LTA selection.
(d) Final preprocessed images after rotation. (e) In the absence of STA/LTA selection, the preprocessed images focus solely on detecting negative amplitude values. The red box emphasizes the comparative clarity of traces, and the brown box illustrates the contrast between vehicle trace and background.}
\label{fig_pre}
\end{figure*}

For data preprocessing, we proposed a workflow for vehicle detection with 1D and 2D DAS processing methods.
The data collected by DAS consists of multi-channel one-dimensional signals, allowing for processing as either separate 1D signals or an integrated 2D image according to detection demand shown in Fig.~\ref{fig_pre}(a).
1D methods often exhibit higher sensitivity to amplitude variation, while 2D methods preserve more spatial positioning information for each channel in real-life settings.
In our 2D image object detection framework, we incorporated 1D event detection methods into the preprocessing stage, aiming to retain the sensitivity of 1D methods to traffic signals.
In addition to standard preprocessing techniques such as detrending, our preprocessing employed model-based filtering and Short-Term Average to Long-Term Average (STA/LTA) in the 1D domain. In the 2D domain, rotation based on speed limit is utilized.

Initially, model-based filtering was applied after detrending and demeaning.
The precise determination of the frequency range is a crucial foundation for achieving optimal filtering performance. 
Regarding real-world vehicles, this information can be acquired through physical modeling.
Vehicles can be considered seismic sources that apply forces continuously at wheel-road contacts.
These force causes strain received by DAS \cite{jousset2018dynamic}. 
In \cite{lindsey2020city}, the entire process from the passage of vehicles to the collection of signals can be described using the Flamant-Boussinesq equation.
Substituting the physical parameters of urban vehicles, the final practical filtering frequency range is approximately 0.1-1 Hz.
Due to the similarity in urban detection tasks, we also maintained this frequency range within our framework shown in Fig.~\ref{fig_pre}(b).

Subsequently, the STA/LTA selection is employed on individual signals to discern potential triggers.
This classical seismic data processing technique computes the ratio of average energy in a short window capturing impulsive events to that in a long window representing ambient noise:

\begin{equation}
\begin{aligned}
\frac{STA}{LTA}(i) = \frac{STA(i)}{LTA(i)}=\frac{ \frac{1}{n_s} \sum_{j=i-n_s}^{i} f^2(j)}{\frac{1}{n_l} \sum_{j=i-n_l}^{i}f^2(j) }   
\label{sta}
\end{aligned}
\end{equation}
where $n_s$ and $n_l$ are the window lengths of STA and LTA, respectively. $f$ represents the amplitude of a signal.
A seismic event is flagged when the STA/LTA ratio exceeds a certain threshold.

This method is crucial for applications such as earthquake early warning and volcanic eruption forecasting due to its simplicity and efficiency. 
\cite{lindsey2020city} extended the STA/LTA algorithm to DAS to detect vehicles by treating their passage as seismic events.

\begin{figure}[!t]
\centering
\includegraphics[width=3in]{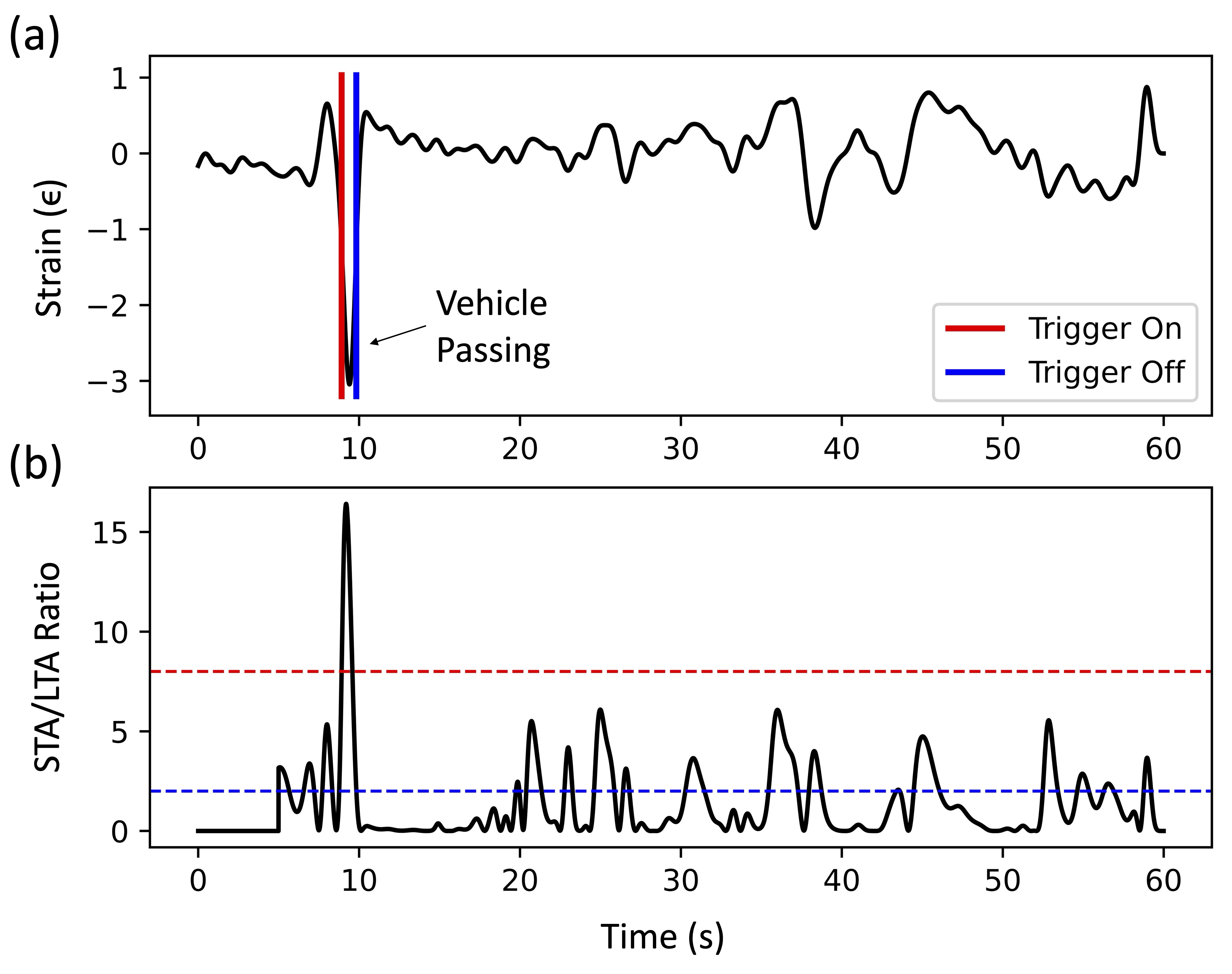}
\caption{Vehicle detection using STA/LTA for a single channel within the Stanford 2 Array (Sand Hill Road Array). (a) Amplitude-time domain; (b) STA/LTA-time domain. The red line signifies the STA/LTA value reaching the initiation threshold, marking the onset of an event, while the blue line indicates the STA/LTA value meeting the termination threshold, denoting the end of the event.}
\label{fig_sta}
\end{figure}

Applying STA/LTA algorithm to each 1D time series, it generated multiple triggers. 
While the STA/LTA algorithm performs well in event detection, using it alone has limitations by ignoring spatial context of nearby DAS channels. So we first applied this algorithm to each channel in the 2D DAS data (Fig.~\ref{fig_sta}). 
Given the negative amplitude of vehicle passages, we only retained the negative-amplitude detections exceeding the  trigger-on threshold as potential vehicle positions. 
These triggers replaced strain time series, integrating into the 2D image and improving feature capture for later trace detection. 
This method produces cleaner images and allows direct application of color domain transformations.
The ablation test for the STA/LTA selection is conducted in Fig.~\ref{fig_pre}(d) and (e).


In the final preprocessing phase, the images were rotated according to the speed limit of Sand Hill Road.
In image-based object detection, detection accuracy significantly decreases when the bounding boxes that delineate individual targets excessively overlap \cite{redmon2016you}.
In this framework, the images were rotated at an appropriate angle to maintain the vertical alignment of the individual traces, thereby alleviating this issue.
To obtain a relatively generalizable angle, it is feasible to calculate it based on the speed limit since most vehicles tend to operate within a speed range below the speed limit.
Sand Hill Road's speed limit is 35 mph, and the rotation angle is calculated to be 55 degrees.
This rotation was uniformly applied to the entire dataset, shown in Fig.~\ref{fig_pre}(d).

\section{Semi-Supervised Vehicle Detection}\label{sec_net}
To address the labeling challenges posed by the high volume and continuous collection of DAS data, our method employed semi-supervised learning for complete labeling and automatic updating.

\begin{figure*}[!t]
\centering
\includegraphics[width=6.7in]{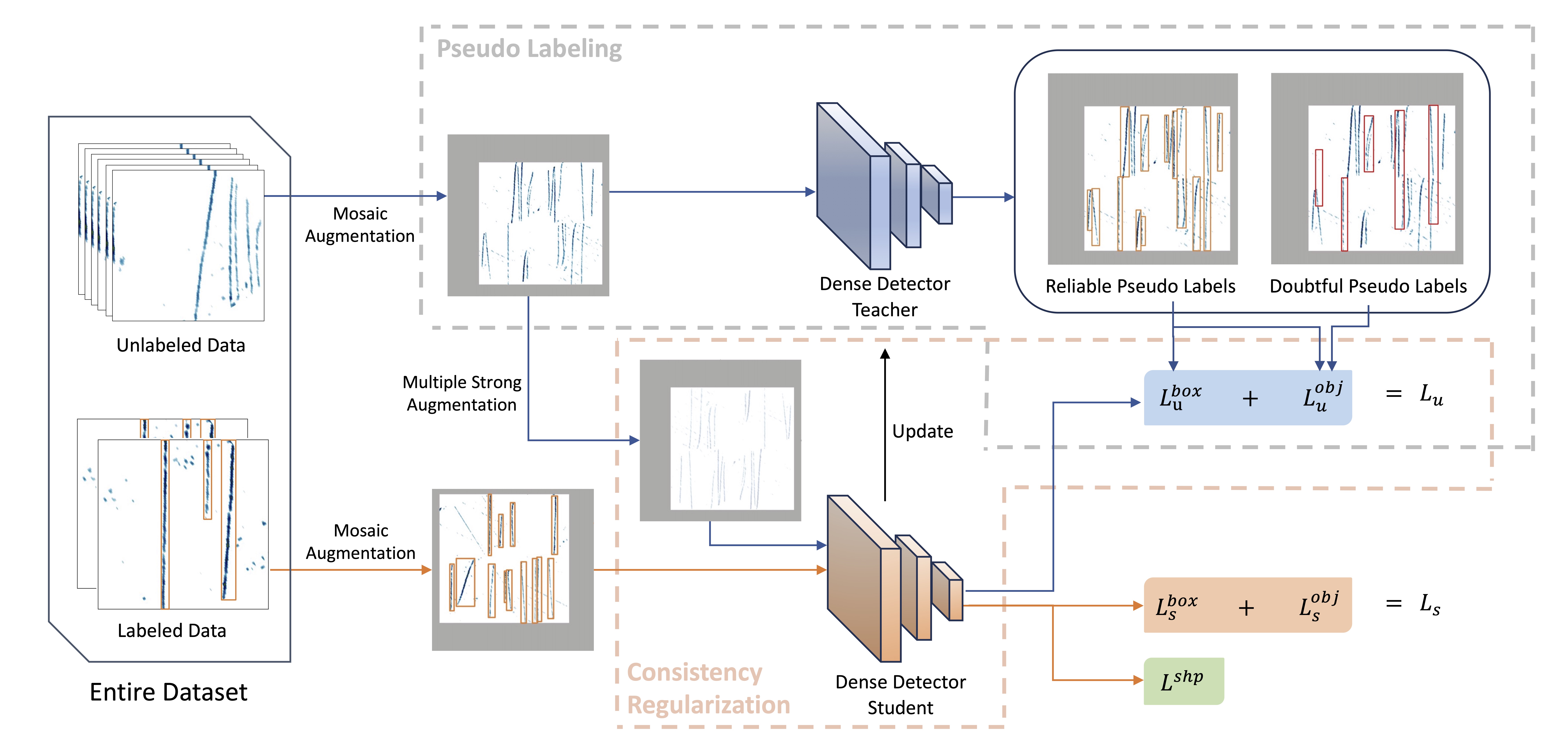}
\caption{Architecture of our vehicle detection model with shape prior loss.
The basic architecture is an Efficient Teacher network \cite{xu2023efficient}.
Blue arrowed lines indicate training with unlabeled data, while orange arrowed lines indicate that with labeled data.}
\label{fig_pipeline}
\end{figure*}

Semi-supervised learning is a machine learning method that leverages a small amount of labeled data and efficiently utilizes the extensive unlabeled data previously discarded \cite{sohn2020simple,berthelot2019mixmatch}.
It is virtually cost-free to enhance model performance and significantly reduces the need for extensive human labels.
This approach is advantageous when obtaining labeled data is expensive or impractical.
Exploiting the training capability of semi-supervised learning with unlabeled data, we have extended this approach to enable continuous model refinement via the acquisition of raw data.

This section introduces the framework proposed for detecting vehicles in real-world traffic flows. It features a semi-supervised algorithm for single-class detection with innovative shape prior loss.

\subsection{Efficient Teacher}
Our base semi-supervised model is an Efficient Teacher network \cite{xu2023efficient}.
In the semi-supervised learning field,  Efficient Teacher distinguishes itself by being the first method to transform the classic fully-supervised YOLO object detector into a semi-supervised framework.
It achieves significant accuracy improvement with considerably fewer labeled samples \cite{redmon2016you,xu2023efficient}.
Moreover, the framework offers a flexible and powerful semi-supervised training architecture that can easily adapt to various real-world applications.
Therefore, this framework can facilitate the transfer and implementation of DAS vehicular detection.

In Efficient Teacher, the basic detector module proposed is called Dense Detector \cite{xu2023efficient}.
This well-designed module comes from a hypothesis validated in \cite{xu2023efficient}, which suggests that increasing the input density can effectively enhance the performance of one-stage anchor-based detectors.
With much denser sampling, Dense Detector achieves higher detection accuracy and better compatibility with semi-supervised learning tasks.

In addition, Efficient Teacher aligns with the crux principles for semi-supervised learning: pseudo labeling and consistency regularization.

\subsection{Network Architecture}
Here we present the network architecture shown in Fig.~\ref{fig_pipeline} while also introducing how pseudo labeling and consistency regularization enable semi-supervised learning.

First, the dataset input to the network is shown on the far left of Fig.~\ref{fig_pipeline}.
There are two input components for training: a small collection of labeled data and a substantially more extensive set of unlabeled training data. In this study, each was derived from 70-channel continuous DAS recordings spanning 60 seconds and subsequently preprocessed into 2D images.
Following the arrow in Fig.~\ref{fig_pipeline}, these data are input into our augmentation.

Subsequently, labeled data build an initial model for semi-supervised learning as represented by orange arrows in Fig.~\ref{fig_pipeline}.
Following image augmentation including scaling and Mosaic processing, they are fed into the student Dense Detector we kept from Efficient Teacher.
The outputs are then utilized to update the Student Dense Detector with supervised learning.
This process implements the classical computation of a loss function followed by backpropagation of gradients to update the model, with the regular standard object detection loss function\cite{redmon2016you}.
The Student Dense Detector updates the Teacher Dense Detector in every iteration, so the Teacher Dense Detector will be updated to an initial model.

Regarding unlabeled data, the first method for semi-supervised learning is pseudo labeling, as shown in the gray dashed box in Fig.~\ref{fig_pipeline}.
It aims to increase the quantity of labeled data.
The initial model, the blue Teacher Dense Detector, inputs a substantial amount of unlabeled data with slight augmentation and changes.
The output labels will be input into the Pseudo Label Assigner.
Two distinct threshold constants of varying magnitudes are employed to assess the reliability of labels output: 1) those exceeding the higher threshold constant $\tau_2$ are marked reliable; 2) any labels falling below the lower threshold $\tau_1$ are disregarded; 3) labels between the two thresholds are considered doubtful.
The ones marked reliable and doubtful will be considered pseudo labels for continued training, effectively broadening the labeled data collection.


Another principle of semi-supervised learning is consistency regularization, which has a distinct purpose and procedure.
It originates from the premise that a robust and accurate model should yield similar outputs for both an original image and its heavily augmented version (e.g., through rotation and color transforms). 
Thus, in our network architecture shown in the pink dashed box in Fig.~\ref{fig_pipeline}, an unlabeled image and its augmented counterpart are input into the Teacher Dense Detector and Student Dense Detector. 
The model is then trained by minimizing the output discrepancies between them.

By integrating these two semi-supervised learning methods with the fully supervised learning approach introduced, we can obtain the complete network structure in Fig.~\ref{fig_pipeline}.
We can optimize and update our model by incorporating the various outputs from this architecture into the loss function.

\subsection{Supervised Loss and Unlabeled Loss}
 
In practice, each randomly selected labeled and unlabeled pair concurrently contributes to gradient backpropagation and model updating with the complete loss function:
\begin{equation}
\begin{aligned}
L=L_s+\lambda_u L_u +\lambda_{sha} L_{sha}
\label{eq_lall}
\end{aligned}
\end{equation}
where $L_s$ represents standard supervised loss, $L_u$ means the loss calculated with unlabeled images and $L_{sha}$ is the shape prior loss we proposed, which is introduced in the next subsection. $\lambda_u $ and $\lambda_{sha}$ are constants to balance each part of the loss.

Regarding standard supervised loss $L_s$, this is the same as the regular standard object detection loss function in YOLO \cite{redmon2016you}:
\begin{equation}
\begin{aligned}
L_{s} &= L_s^{box}+L_s^{o b j} \\
&=\sum_{h, w}\left(C I o U\left(Y_{(h, w)}^{box}, X _{(h, w)}^{box}\right)\right.  \left.+C E\left(Y_{(h, w)}^{o b j}, X_{(h, w)}^{o b j}\right)\right) 
\label{eq_Ls}
\end{aligned}
\end{equation}
where CIoU denotes the complete intersection over union value between the predicted box $X_{(h, w)}^{box}$ and the labeled $Y_{(h, w)}^{box}$ at location $(h,w)$ on the feature map, while CE denotes the cross-entropy loss function comparing the predicted confidence of object presence $X_{(h, w)}^{o b j}$ to the labeled counterpart $Y_{(h, w)}^{o b j}$. 
Since this is a single-class detection for vehicular traces, the class loss has been discarded.

The loss calculated with unlabeled images includes both pseudo labeling and consistency regularization at the same time.
As shown in Fig.~\ref{fig_pipeline}, pseudo labeling and consistency regularization ultimately converge to $L_u$.
The detection results from the Student Dense Detector are utilized in conjunction with the outputs from the Pseudo Label Assigner to compute the unlabeled loss function.
The outputs from the Pseudo Label Assigner comprise both reliable and doubtful pseudo labels.
Pseudo labels marked as reliable are used in the computation of the standard supervised loss function as equation (\ref{eq_Ls}).
Regarding the doubtful labels, a soft loss is computed based on the magnitude of their objectness score.
Finally, the complete unlabeled loss formula as in \cite{xu2023efficient} is:

\begin{equation}
\begin{aligned}
L_u=L_u^{box}+L_u^{o b j}
\label{eq_lu}
\end{aligned}
\end{equation}
where
\begin{equation}
\begin{aligned}
L_u^{box}=\sum_{h, w}\left(\mathbbm{1}_{\left\{p_{(h, w)}>=\tau_2 \text { or }  \hat{obj} _{(h, w)}>0.99\right\}}\times\right.\\
\left.\operatorname{CIoU}\left(X_{(h, w)}^{box}, \hat{Y}_{(h, w)}^{box}\right) \right)
\label{Lureg}
\end{aligned}
\end{equation}

\begin{equation}
\begin{aligned}
L_u^{o b j} & =\sum_{h, w}\left(\mathbbm{1}_{\left\{p_{(h, w)}<=\tau_1\right\}} C E\left(X_{(h, w)}^{o b j}, \mathbf{0}\right)\right. \\
& \left.+\mathbbm{1}_{\left\{p_{(h, w)}>=\tau_2\right\}} C E\left(X_{(h, w)}^{o b j}, \hat{Y}_{(h, w)}^{o b j}\right)\right) \\
& \left.+\mathbbm{1}_{\left\{\tau_1<p_{(h, w)}<\tau_2\right\}} C E\left(X_{(h, w)}^{o b j}, o \hat{b} j_{(h, w)}\right)\right)
\label{Luobj}
\end{aligned}
\end{equation}
$X_{(h, w)}^{o b j}$ and $X_{(h, w)}^{box}$ are outputs of student Dense Detector as equation (\ref{eq_Ls}). $\hat{Y}_{(h, w)}^{obj}$ and $\hat{Y}_{(h, w)}^{box}$ represent the presence and location score calculated by Pseudo Label Assigner on the feature map, while $\hat{obj}$ indicates the objectness score of pseudo label at location $(h,w)$. $\mathbbm{1}$ is an indicator function that is equal to $1$ if certain conditions are met and $0$ otherwise. $\tau_1$ and $\tau_2$ are threshold constants of Pseudo Label Assigner.

With the loss calculated in equation (\ref{eq_lall}), the student Dense Detector is continuously updated via loss gradient descent. 
Simultaneously, the teacher Dense Detector is enhanced through Exponential Moving Average (EMA) updates. 
This process forms a learning cycle that iteratively advances both detectors, achieving semi-supervised learning.

\subsection{Shape Prior Loss}

\begin{figure}[!t]
\centering
\includegraphics[width=3.1in]{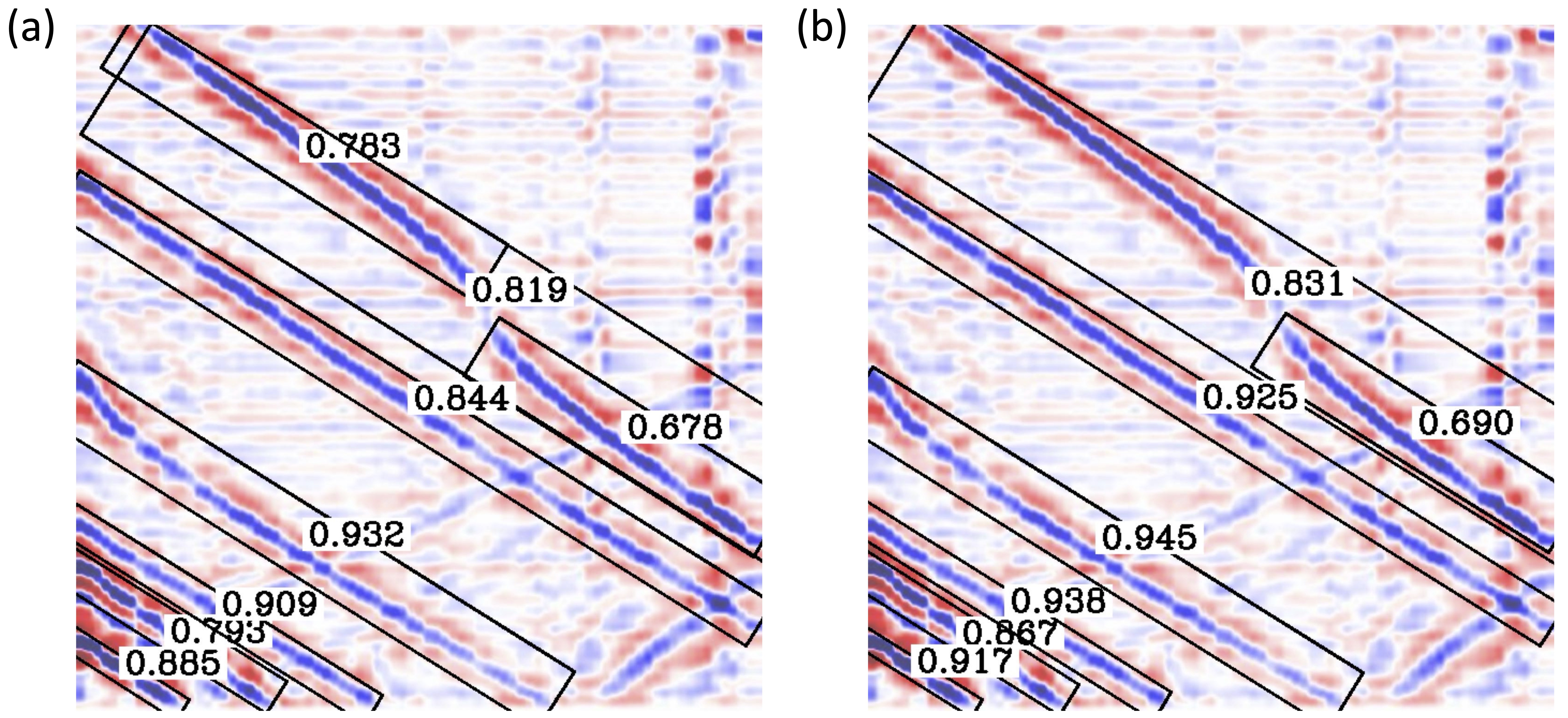}
\caption{Apparent false detection in the top right corner erroneously identifies one trace as even three in (a) and two in (b).
The black numbers represent the confidence scores, which indicate the model's certainty regarding the accuracy of each prediction.}
\label{fig_lshape}
\end{figure}

A scarcity of labeled data inevitably results in erroneous detections.
For example, due to the variations in vehicle speed and intermittent signal loss caused by noise, the vehicle trace often appears as a discontinuous curve in images.
This leads models trained on small-scale datasets to erroneously identify a single trace as multiple traces with overlapping boxes, as shown in Fig.~\ref{fig_lshape}.
When direct training on datasets yields unsatisfactory results, a practical method to address such erroneous detections becomes critical.

\begin{figure}[!t]
\centering
\includegraphics[width=2.2in]{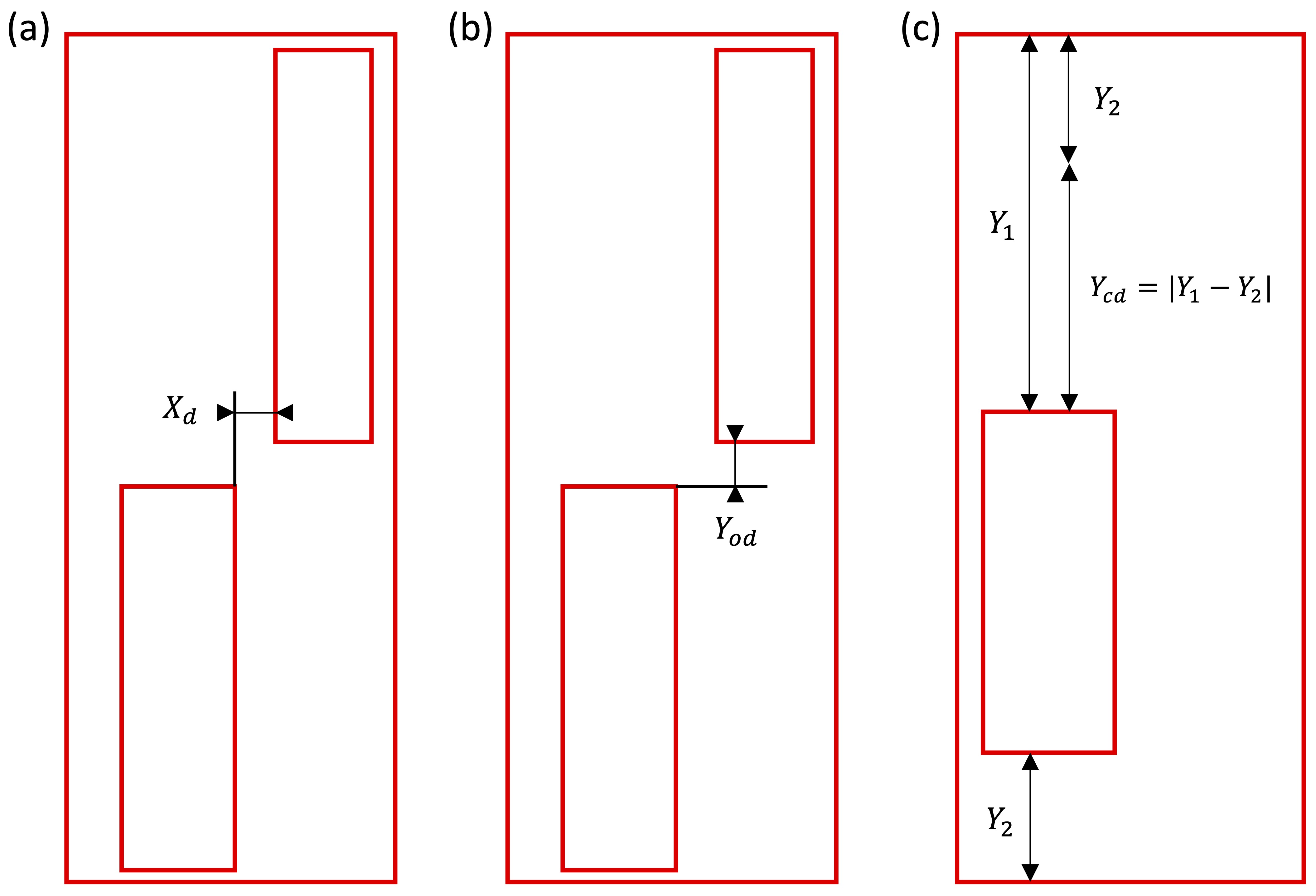}
\caption{Illustration of the three distances in shape prior loss. (a) Minimum vertical edge distance for opposite and corresponding box edges $X_d$. (b) Shortest horizontal edge distance for opposite box sides $Y_{od}$ (c) Absolute difference in nearest same-orientation horizontal edge distance $Y_{cd}$.}
\label{fig_lshafig}
\end{figure}

\begin{figure*}[!t]
\centering
\includegraphics[width=7.1in]{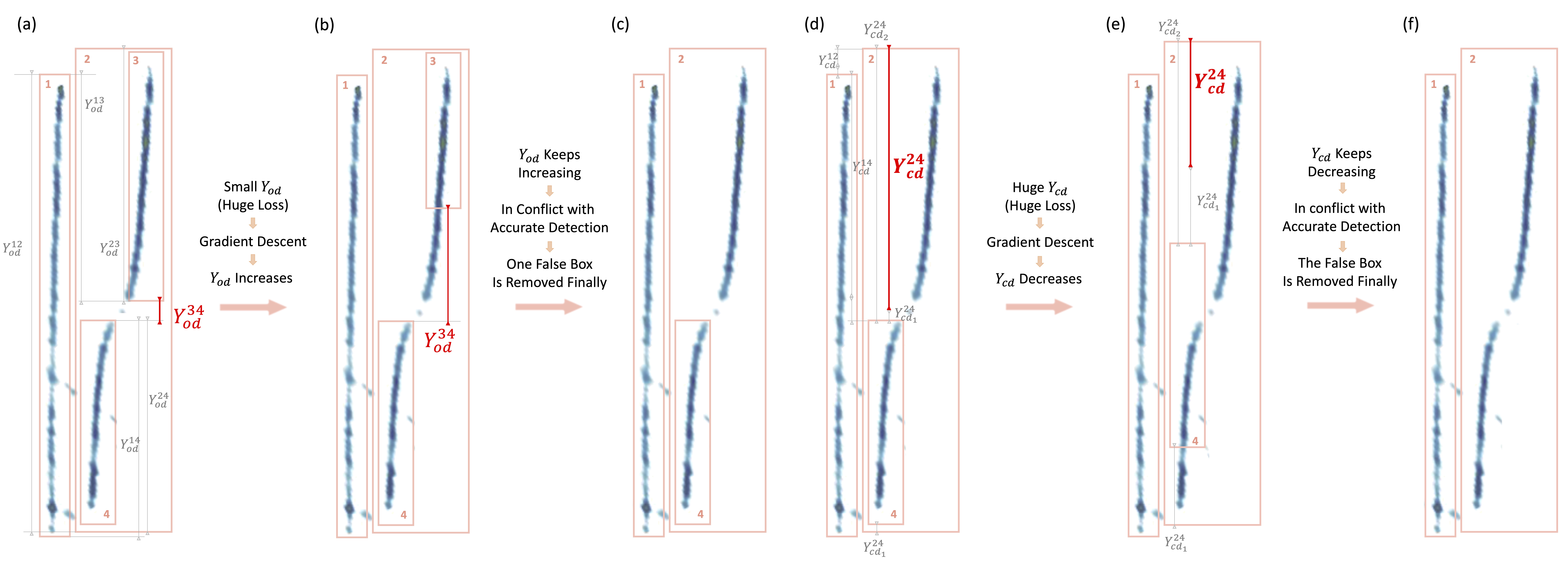}
\caption{Demonstration of shape prior loss in eliminating false detection boxes. 
(a)-(c) show the impact of $L_{Y_{od}}$ with multiple false detections; (d)-(f) show the impact of $L_{Y_{cd}}$ with a single remaining false detection.}
\label{fig_yodloss}
\end{figure*}

Imparting knowledge to a model does not necessarily mean learning freely from the dataset.
Instead, it can be achieved by incorporating prior knowledge for targeted learning \cite{el2021high}.
With prior information, it can compensate for limited data by guiding the learning process and improving generalizability.
In this study, we incorporated simple shape priors into the loss function and treated the erroneous detection (e.g., the boxes in the top right of Fig.~\ref{fig_lshape}) as a penalizing loss term to avoid these predictions:

\begin{equation}
\begin{aligned}
L_{sha} 
&= \lambda_{od}L_{Y_{od}}+ \lambda_{cd}L_{Y_{cd}}\\
\label{eq_Lsha}
\end{aligned}
\end{equation}
\begin{equation}
\begin{aligned}
L_{Y_{od}}
&= (1-Y_{od})\times detach(1-X_d)\\
L_{Y_{cd}}
&= Y_{cd}\times detach(1-X_d)\\
\end{aligned}
\end{equation}
where $L_{sha}$ means shape prior loss. $X_d$ is a pairwise distance matrix between the boxes and the nearest vertical edges as in Fig.~\ref{fig_lshafig}(a). $Y_{od}$ indicates the distances between the nearest opposite (e.g. top-bottom) horizontal edges of two boxes as in Fig.~\ref{fig_lshafig}(b). $Y_{cd}$ captures the difference in distances between the furthest horizontal edges and nearest horizontal edges of the same side (e.g. top-top or bottom-bottom) of two boxes as in Fig.~\ref{fig_lshafig}(c). 
Detach means the variable is removed from the computation graph and does not change during gradient descent.

An intuitive interpretation of shape prior loss is that given our prior knowledge that the traces are characterized by almost non-intersecting vertical rectangles, many boxes that do not conform to these criteria can be considered erroneous detections.
Consequently, we can add this insight into our loss function to mitigate such inaccuracies.
Specifically, the loss peaks when two boxes are horizontally close, with either their same-side or opposite-side edges nearing each other.
It is aligned with the common false detection observed in Fig.~\ref{fig_lshape}. 

Each term indicates different types of distances. 
$1-X_d$ is multiplied with other terms to introduce horizontal distances, ensuring near boxes yield a greater loss.
As a penalty factor, the value should increase as the distances decrease, hence the use of $1-X_d$ instead of $X_d$ itself.
In addition, since the horizontal distances do not need to change, $1-X_d$ does not participate in gradient descent and is detached from the computational graph.
Regarding $1-Y_{od}$, it ensures that the model avoids two horizontally adjacent boxes from being too close in the vertical direction.
Thus, it effectively prevents the erroneous scenario depicted in Fig.~\ref{fig_lshape}(a).
Similarly, $Y_{od}$ is a value where a larger loss function indicates better model performance.
Therefore, $1-Y_{od}$ is utilized but $Y_{od}$.
The presence of $Y_{cd}$ is to prevent a scenario during a model update where one small box vanishes while another incorrect small box remains by imposing a high loss when a small box is near the edge of a larger box as in Fig.~\ref{fig_lshape}(b).

Specifically, Fig.~\ref{fig_yodloss} illustrates how shape prior loss incrementally removes false detections through gradient descent. 
In Fig.~\ref{fig_yodloss}(a), when multiple false detections first appear, ${Y_{od}}$ is calculated between each pair of boxes. 
${Y_{od}}$ between false detections (e.g., $Y_{od}^{34}$ in Fig.~\ref{fig_yodloss}(a)) is obviously small, indicating a large loss in $1-{Y_{od}}$. 
This value is amplified again through exponential scaling. 
Reducing this loss increases ${Y_{od}}$, leading to Fig.~\ref{fig_yodloss}(b). 
Since $X_d$ is detached and not part of the gradient computation graph, it will not be modified to reduce the loss.
After the box position is adjusted based on the shape prior loss, the false box will not remain with a large ${Y_{od}}$. 
Since there is no complete or correct detection within the box, the loss in Eq.~\ref{eq_Ls} and Eq.~\ref{eq_lu} increases. 
The gradient direction that minimizes all losses eventually leads to removing the box.
Repeated iterations eventually remove the erroneous box in Fig.~\ref{fig_yodloss}(c) as the loss decreases.
Similarly, in Fig.~\ref{fig_yodloss}(d), ${Y_{cd}}$ is calculated between each pair of boxes. The large ${Y_{cd}}$ between a false detection and the correct result indicates a large loss. Reducing this loss requires decreasing ${Y_{cd}}$, leading to Fig.~\ref{fig_yodloss}(e). Eventually, the erroneous box will be removed as the model is updated in Fig.~\ref{fig_yodloss}(f).
It is worth noting that the two losses are calculated and applied to gradient descent simultaneously.

A detailed description of shape prior loss calculation is provided in Algorithm \ref{alg}.
In practical computation, we employed extensive matrix representation and operations in a single image to fully leverage GPUs' substantial advantages in tensor computation speed.
For a given image, each distance term between any two boxes is represented using an element of a n*n matrix, where n is the number of boxes. 
The shape prior loss in matrix form between the two boxes is then calculated.
Finally, these are summed to obtain a scalar shape prior loss for gradient descent.
   
\begin{breakablealgorithm}
	\caption{Shape Prior Loss Calculation  with Bounding Boxes Predicted}
	\label{alg}
	\begin{algorithmic}[1]
        \REQUIRE
			prediction matrices of an image with specific bounding box width $W$, height $H$, and center coordinates $(X,Y)$, the number of bounding boxes $n$
		\ENSURE
			shape prior loss $L_{sha}$ for a single image

         
         \STATE{// Pairwise distance calculation with triangular matrix representation for effective computation on GPUs}
             
             

         \FOR{$i \gets 1$ to $n$}
    \FOR{$j \gets 1$ to $i$}
        \STATE{// For the boxes and the nearest vertical edges }
        \STATE{$X_i^{right} \gets X[i] + W[i]/2$}
        \STATE{$X_i^{left} \gets X[i] - W[i]/2$}
        \STATE{$X_j^{right} \gets X[j] + W[j]/2$}
        \STATE{$X_j^{left} \gets X[j] - W[j]/2$}
        \STATE{$X_{od}[i, j] \gets \min(X_i^{right} - X_j^{left}, X_i^{left} - X_j^{right})$}
        \STATE{$X_{cd}[i, j] \gets \min(X_i^{right} - X_j^{right}, X_i^{left} - X_j^{left})$}
        \STATE{$X_d[i, j] \gets \min(X_{od}[i, j], X_{cd}[i, j])$}
        
        \STATE{// For opposite horizontal edges}
        \STATE{$Y_i^{top} \gets Y[i] + H[i]/2$}
        \STATE{$Y_i^{bottom} \gets Y[i] - H[i]/2$}
        \STATE{$Y_j^{top} \gets Y[j] + H[j]/2$}
        \STATE{$Y_j^{bottom} \gets Y[j] - H[j]/2$}
        \STATE{$Y_{od}[i, j] \gets \min(Y_i^{top} - Y_j^{bottom}, Y_i^{bottom} - Y_j^{top})$}
        
        \STATE{// For corresponding horizontal edges}
        \STATE{$Y_{cd}[i, j] \gets \left| (Y_i^{top} - Y_j^{top}) - (Y_i^{bottom} - Y_j^{bottom}) \right|$}
    \ENDFOR
\ENDFOR

		 \STATE{// Normalization and exponential amplification}
         \STATE {$Z \gets \exp(Z / \max(Z))$ for $Z \in \{X_d, Y_{cd}, Y_{od}\}$}

		 \STATE{// A smaller distance generates a larger loss for $X_d$ and $Y_{od}$}
\FOR{$Z \in \{X_d, Y_{od}\}$}
    \STATE{$L_Z \gets 1 - Z/\max(Z)$}
\ENDFOR
\STATE{$L_{Y_{cd}} \gets Y_{cd}/\max(Y_{cd})$}
		 
		 \STATE {$L_Z \gets L_Z + \lambda_Z \times \text{mean}(L_Z)$ for $Z \in \{Y_{od}, Y_{cd}\}$}

		 \STATE {// Shape prior loss matrix calculation}
		 \STATE {$L_{Msha} \gets L_{Y_{od}}\times \text{detach}(L_{X_d}) + L_{Y_{cd}}\times \text{detach}(L_{X_d})$}
		 \STATE {// Final shape prior loss scholar}
		 \STATE {$L_{sha} \gets \sum(L_{Msha})/n^2$}
         
	\end{algorithmic}
 \end{breakablealgorithm}

\section{Experimental Results}\label{sec_exp}
In this section, we evaluated our model's capability for vehicle monitoring compared with other mainstream methods.
We also conducted experiments on automatic updating, validating various updating strategies.

\renewcommand{\dblfloatpagefraction}{.9}
\begin{figure*}[!t]
\centering
\includegraphics[width=5.6in]{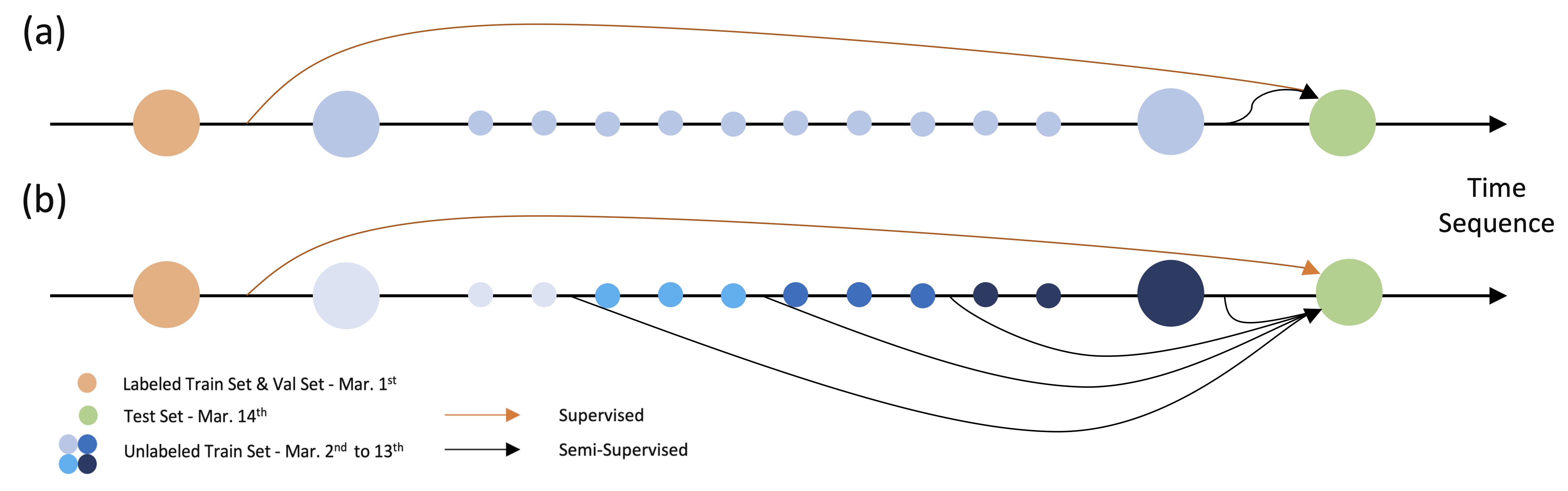}
\caption{Schematic Diagram of Experimental Tasks. (a) Detection performance is tested using all available unlabeled data in a single instance. (b) The model is continually updated with newly acquired unlabeled data to test its performance in simulating real-world conditions by exemplifying a three-day interval. 
In both tasks, data labeled on the first day are employed for fully supervised training as a basis for comparison.}
\label{fig_task}
\end{figure*}

\subsection{Detection Performance}
\subsubsection{Experimental Setting}
To assess the model's performance, we first trained and validated it on small labeled datasets of varying sizes and the same unlabeled dataset.
After that, we tested on a separate test set to obtain metrics.
In comparative experiments, the frequently employed line detection method Hough Transform \cite{catalano2021automatic,corera2023long,litzenberger2021seamless}, the original supervised YOLO model \cite{redmon2016you,ye2023traffic}, and Efficient Teacher \cite{xu2023efficient} without shape prior loss we proposed were evaluated under the same setting.

In our study, the dataset was not split by the conventional random allocation across training, validation, and test sets.
Our split aligned with the operational workflow of the real-world detection task: initial data collected was labeled and utilized to train a full-supervised model, followed by iterative model refinement using subsequently gathered unlabeled data over time. 
Consequently, data spanning 14 days was split by the first 13 days for training and validation, while data from the final day served as the test set. 
Of the initial 13 days, only the first day was labeled, leaving the data from the second to the thirteenth day unlabeled, shown in Fig.~\ref{fig_task}(a).

To thoroughly test the limit of our model with fewer labeled data, we worked on a series of experiments that reduced the labeled dataset by half at each test.
The scale of the dataset is in Table \ref{tab_scale}.

\begin{figure*}[!t]
\centering
\includegraphics[width=6.5in]{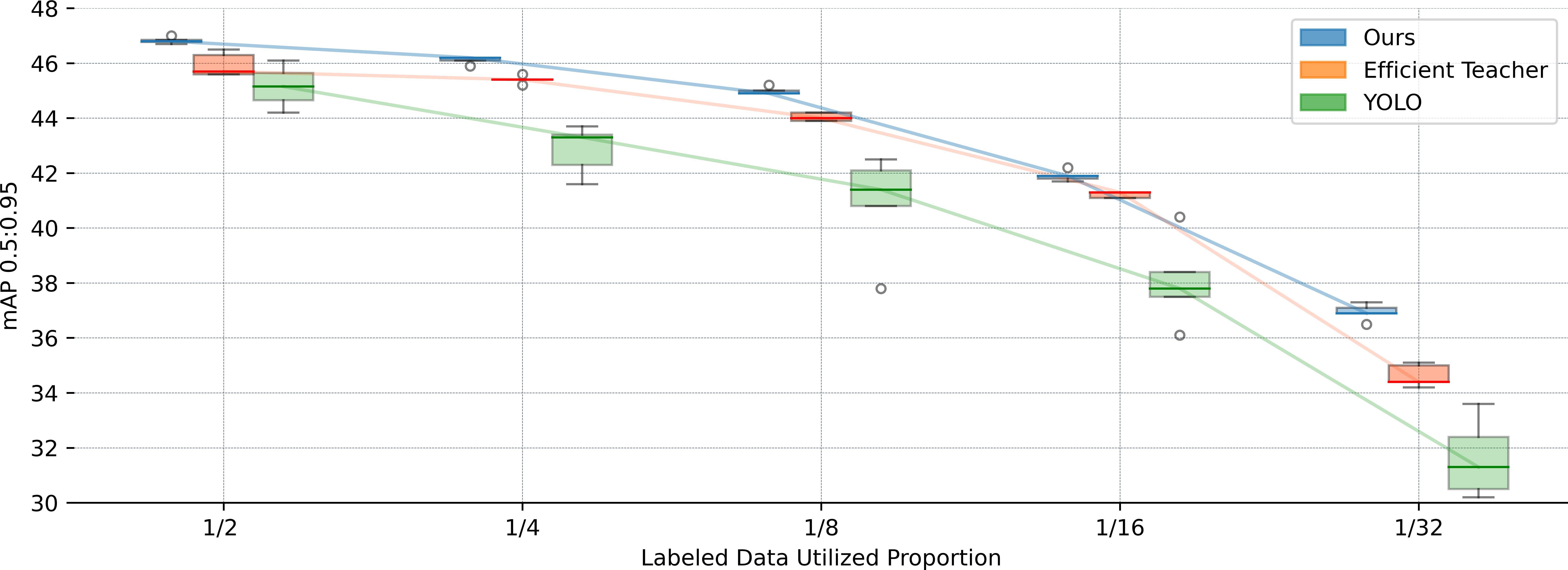}
\caption{The central rectangle of each plot spans from the first to the third quartile, a line inside marks the median, and whiskers extend to the minimum and maximum values, with outliers plotted as individual points.}
\label{fig_mapre}
\end{figure*}

\begin{table*}[!t]
\caption{Data Split\label{tab:table1}}
   \centering

     \begin{tabular}{|c|c|c|c|c|c|c|c|c|}
     \hline
     \multirow{3}[6]{*}{Set} & \multicolumn{7}{c|}{Labeled Data}                          & \multirow{3}[6]{*}{Unlabeled Data} \bigstrut\\
\cline{2-8}           & \multicolumn{5}{c|}{Train Set}        & \multirow{2}[4]{*}{Val Set} & \multirow{2}[4]{*}{Test Set} &  \bigstrut\\
\cline{2-6}           & 1/2   & 1/4   & 1/8   & 1/16  & 1/32  &       &       &  \bigstrut\\
     \hline
     Number of Samples & 551   & 276   & 138   & 69    & 35    &   114    & 590   & 13818 \bigstrut\\
     \hline
     \end{tabular}%
   \label{tab_scale}%
 \end{table*}%

\subsubsection{Metric}
The performance of our model is evaluated by mean Average Precision (mAP), a classic metric in object detection algorithms.
It differs from conventional accuracy by involving both precision (the proportion of true positives among all detected items) and recall (the proportion of true positives detected among all actual positives), 

\begin{equation}
\begin{aligned}
\text{Precision} = \frac{TP}{TP + FP}
\label{tp}
\end{aligned}
\end{equation}
\begin{equation}
\begin{aligned}
\text{Recall} = \frac{TP}{TP + FN}
\label{tp}
\end{aligned}
\end{equation}
where True Positives (TP), True Negatives (TN), False Positives (FP), and False Negatives (FN) are the generally four classes of model detection results. 
Their definitions are as follows:
\begin{itemize}
    \item TP: the number of positive samples correctly detected as positive samples.
    \item TN: the number of negative samples correctly detected as negative samples.
    \item FP: the number of negative samples incorrectly detected as positive.
    \item FN: the number of positive samples incorrectly detected as negative.
\end{itemize}
where positive sample means the presence of objects and negative means background.

mAP is eventually calculated by taking the area under the precision-recall curve, thus offering a more robust measure of an algorithm's performance. 
In the context of single-class detection, the formula below represents mAP, which is equivalent to AP:

\begin{equation}
\begin{aligned}
mAP = AP_i = \int_{0}^{1} p(r) \, dr
\label{map}
\end{aligned}
\end{equation}
where $p(r)$ is the precision at a given recall $r$. 

The metric mAP is differentiated into mAP 0.5 and mAP 0.5:0.95.
The former evaluates detection precision at a single Intersection over Union (IoU) threshold of 0.5, while the latter averages precision across IoU thresholds from 0.5 to 0.95 in 0.05 increments.
Therefore, mAP 0.5:0.95 provides a more comprehensive assessment across a range of localization accuracies. For these reasons, we selected mAP 0.5:0.95 as our metric.

\subsubsection{Implementation Detail}
To ensure the experiments' reliability and mitigate the effects of randomness in training, we repeated the experiments five times across various labeled data scales. 
We analyzed the results using box plots.
Box plots visually present the distributions of results by presenting their five-number summaries: minimum, first quartile, median, third quartile, and maximum.  
The weights for loss function are as follows: the standard supervised loss $\lambda_s$ weights $1$, the loss from unlabeled data weights $\lambda_u = 3$, and the shape prior loss weights $\lambda_{sha} = 0.3$. 
In the shape prior loss equation, the weights are $\lambda_{od} = 6$ for the horizontal opposite edge loss and $\lambda_{cd} = 1$ for the horizontal corresponding edge loss.

\subsubsection{Results}
Statistical data of comparative experimental results are displayed in Fig.~\ref{fig_mapre}.
Compared with the semi-supervised Efficient Teacher and the supervised YOLO model, our method outperformed on the metric of mAP 0.5:0.95 for any given volume of labeled data.
This result reflects improved detection accuracy and reduced false positives of our model.
The improved mAP compared with YOLO shows the benefit of the semi-supervised strategy using unlabeled data to augment model performance. 
Our model's results compared to Efficient Teacher demonstrate the success of the shape prior loss without the need for additional data.
Even when operating with just a quarter of the labeled data, our model simultaneously surpassed both the Efficient Teacher and YOLO under half-labeled data training, effectively delivering superior performance with half the labeled resources employed by our counterparts.

From a trend perspective, there is a decline in mAP consistent with the decreasing volume of labeled data. 
However, our method demonstrated the least steep decline in performance.
As the amount of labeled data decreases, our model shows a clearer accuracy advantage over other models.
Furthermore, our model exhibited enhanced robustness and resistance to randomness across different amounts of labeled data, as evidenced by the thinner boxes and fewer outliers in the box plots compared to other methods.

\begin{figure*}[!t]
\centering
\includegraphics[width=6in]{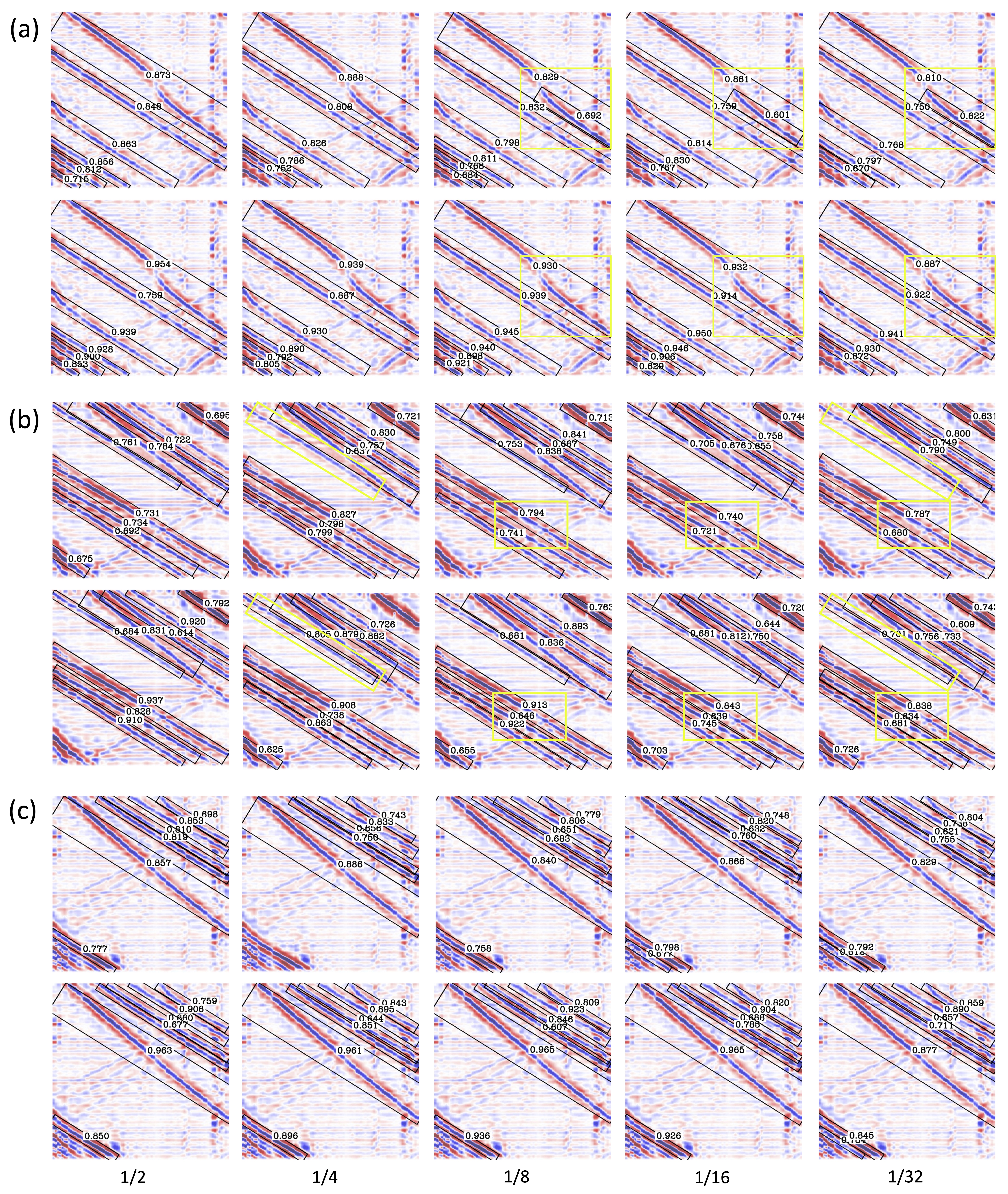}
\caption{Inference instances. 
Within each panel, the upper tier delineates the efficacy of the supervised model, whereas the lower tier reveals the performance of our novel unsupervised approach. Adjacent to the figure, the denoted volume of labeled data diminishes by half stepwise from left to right. Panel (a) delineates the model's performance across the entire curve, while panels (b) and (c) particularly highlight its robustness during peak events. Strategic highlighting with yellow boxes accentuates areas where our model enhances detection confidence and markedly minimizes instances of undetected events.
}
\label{fig_infer}
\end{figure*}

Clearer examples of the inference are shown in Fig.~\ref{fig_infer}. 
The top row of each panel shows the fully-supervised model, while the second row shows our semi-supervised model using the same amount of labeled data. Each column uses successively halved amounts of labeled data from left to right. 
Overall, our model showed higher confidence in trace detection and reduced false negatives compared to supervised methods that do not use unsupervised data.

\begin{figure}[!t]
\centering
\includegraphics[width=3in]{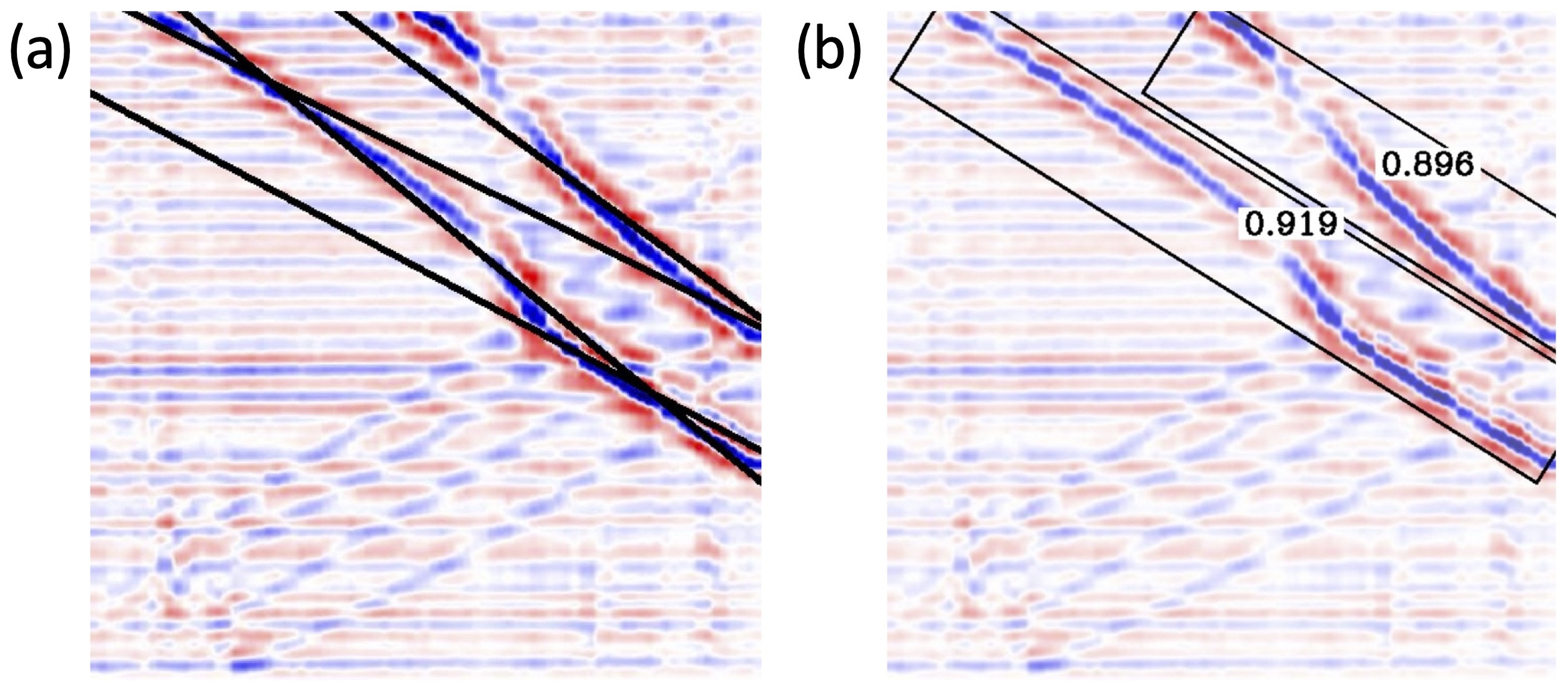}
\caption{Illustrative comparison between (a) Hough Transform and (b) our model of variable-speed curve detection.}
\label{fig_inferhough}
\end{figure}

\begin{figure}[!t]
\centering
\includegraphics[width=3in]{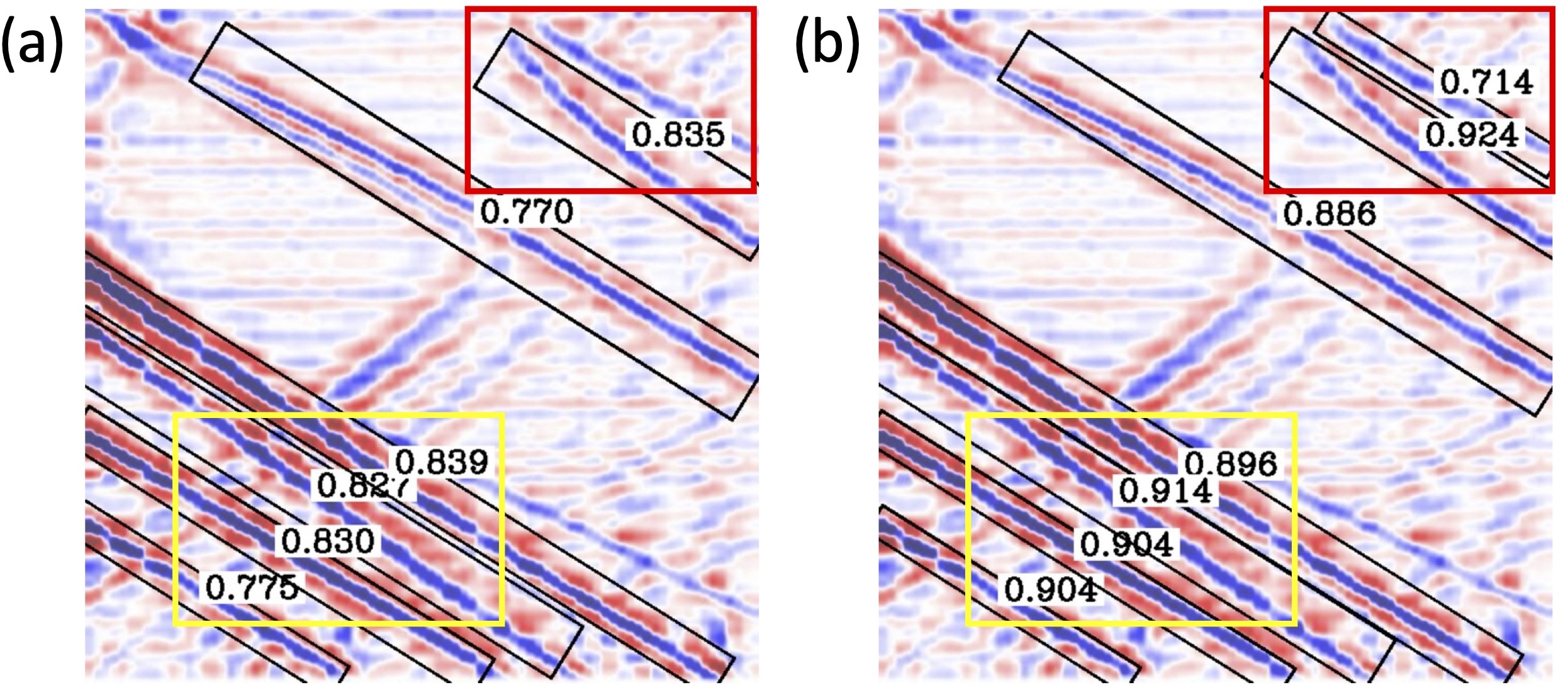}
\caption{Inference examples of (a) supervised model employing 1/2 of the labeled data compared to (b) our model utilizing only 1/32 of the labeled data.}
\label{fig_inferone}
\end{figure}

Our approach accurately detects curves created by vehicles moving at different speeds, as shown in Fig.~\ref{fig_infer}(a). In contrast, the supervised YOLO model struggles with this when the amount of labeled images is reduced to less than 1/8.
Additionally, the curves show details that traditional Hough Transform line detection methods miss, as shown in Fig.~\ref{fig_inferhough}.
Our model also performs well during peak traffic flow, accurately capturing high-density linear and curve features, as seen in Fig.~\ref{fig_infer}(b) and (c).
Using only 1/32 of the labeled data supplemented with unlabeled data, our method produced some results that outperformed supervised models trained with 1/2 of the labeled data, as shown in Fig.~\ref{fig_inferone}.
For tasks where inference time is of paramount importance, our model has real-time monitoring capabilities, processing 60-second DAS data segments in approximately 16 ms—well within the data acquisition duration.

\subsection{Automatic Update}
\subsubsection{Experimental Setting}
To evaluate the effectiveness of automatic updates and identify the best strategy, we tested various training methods that reflect real-world conditions. This involved updating with newly acquired data, either with or without historical data, on a three- or six-day cycle, as shown in Fig.~\ref{fig_task}(b).
The outcome of training with all unlabeled data in a single shot served as a comparison.
Each strategy conducted five trials to mitigate randomness and test robustness.
\subsubsection{Metric}
The metric mAP 0.5:0.95, which comprehensively gauges accuracy, including false detections, was consistently applied in this experiment.

\subsubsection{Implementation Detail}
The automatic update experiment was also conducted on a single NVIDIA RTX A4000 with 16 GB of memory, an initial learning rate of 0.001, and a batch size 12.
The weights are loss function remains the same.

\begin{figure*}[!t]
\centering
\includegraphics[width=6.4in]{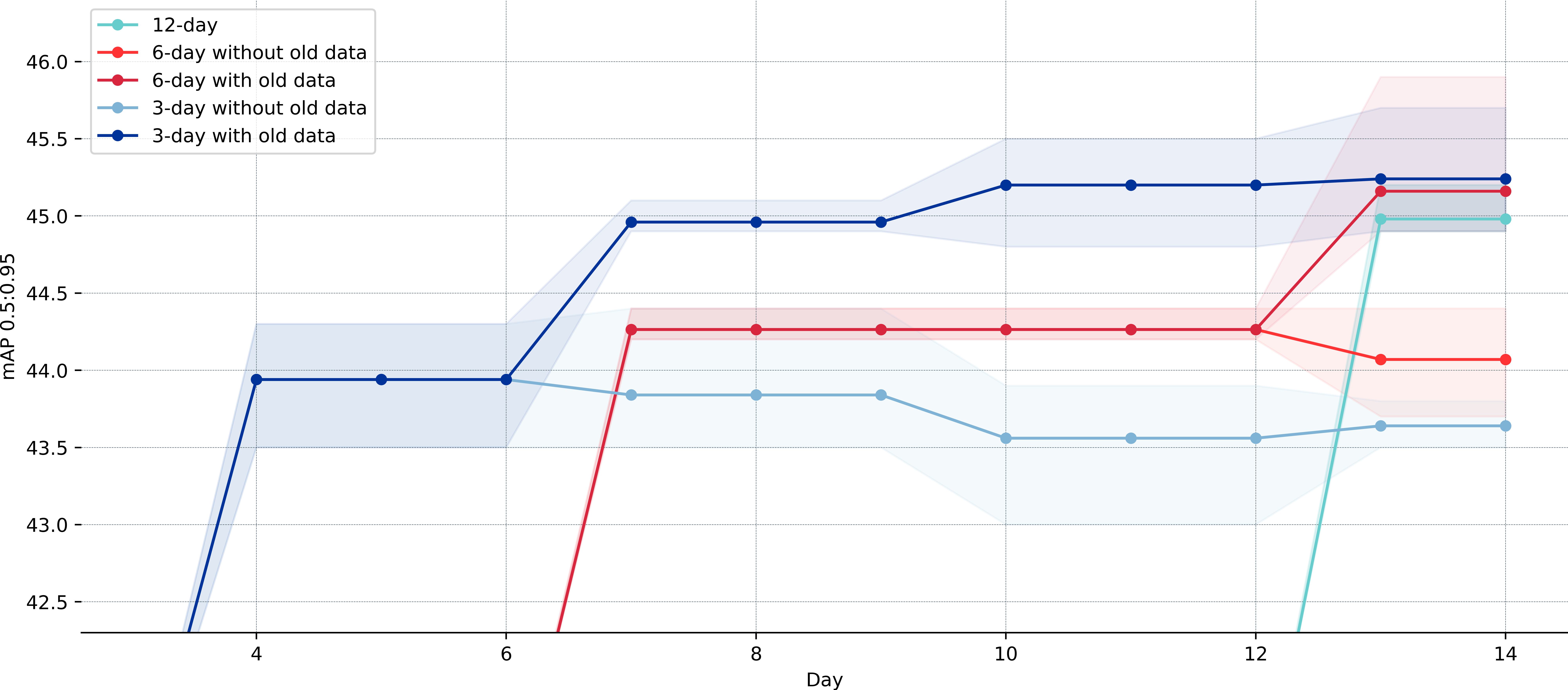}
\caption{Results of various update strategies. In the line graph, points represent the means, while the shaded areas indicate the extremities from five repeated trainings.
The starting point of the lines represents the mAP of the supervised model, trained with the first day's labeled data, averaging 40.92 over five experiments.
}
\label{fig_update}
\end{figure*}

\subsubsection{Results}
The results of various strategies are shown in a line graph in Fig.~\ref{fig_update}. The data clearly indicates that updating the model with new data every three or six days initially improves monitoring efficacy, but further updates can unexpectedly reduce performance.
After 12 days of incremental updates, the efficacy significantly falls short of the strategy that uses a single update with all data from the 12 days. In contrast, combining new data with existing datasets for joint training consistently increased accuracy, whether updates occurred every three or six days.
The gradual model updates using these two strategies over a 12-day period performed better on the test set than a one-time use of all unlabeled data. This finding is quite different from our initial expectations.
Notably, the multiple training iterations theoretically led to overfitting, resulting in diminished performance, even though both strategies used the entire set of unlabeled data in the final update. Our reasoning and conjectures about these results will be further discussed in Section~\ref{sec_dis}.

Both three-day and six-day update intervals have advantages in selecting the best strategy. With only 12 days of unsupervised data, accuracy for the three-day interval began to level off, while the six-day interval continued to improve.
This suggests that continuing to train at six-day intervals may further enhance detection performance beyond what is achievable with three-day intervals. However, the six-day interval results in lower short-term detection performance than the three-day strategy, as shown in the period from day 7 to 12 in Fig.~\ref{fig_update}.
Therefore, we suggest choosing an update strategy based on the duration of the application cycle. 
Specifically, when combining newly collected data with existing data for training, longer intervals are better for extended time duration, while more frequent updates are preferable for shorter durations.

\section{Discussion and Future Work}\label{sec_dis}

In this section, we discuss some of the experimental outcomes, identify areas where our framework requires improvement, and provide directions for future work.

In experiments involving autonomous updates, the outcomes diverged significantly from our expectations. Unsurprisingly, training exclusively with newly collected data leads to a gradual decline in performance, as batch-wise training often results in incorrect gradient descent directions and convergence to local minima.
Under the strategy that uses both new and existing data, the final automatic update also employed the entire dataset, which might have led to overfitting on the older data. Even when mAP values converge, a model trained in a single session is usually expected to perform slightly better.
However, contrary to our expectations, the detection performance of iterative updates surpassed that achieved by a single training session with the complete dataset.


Our interpretation is the noise or mislabeled samples inherent in the dataset. The input data are entirely unlabeled, and there is no guarantee that the pseudo labels assigned by the model are completely accurate, which introduces significant noise.
Adding all the unlabeled data at once affects the model's learning process because the data distribution is unclear and noisy. Therefore, it is essential to gradually incorporate these complex unlabeled data, allowing the model to adapt and engage in incremental learning progressively.
This approach aligns with the challenges that Curriculum Learning aims to address \cite{bengio2009curriculum}.

The reduced performance observed when updates were conducted only with newly added data, as shown by the descending line in Fig.\ref{fig_update}, can be attributed to the failure in finding the accurate gradient descent towards the global optimum of the loss function.
The curve representing three-day interval updates in Fig.\ref{fig_update}, which shows a slight increase during the last three days compared to declines in the initial updates, further supports our hypothesis.
Therefore, we conclude that both the gradient towards the global optimum and the temporal distance between the training data and the inference data are significant, with the former being more critical. 
These findings can inform the development of future update strategies.

Our framework still has space for improvement. Given the limited sample size and the variability of vehicle trajectories in real-world environments, our model may need to improve in specific scenarios, such as extended periods of parallel vehicle movement. 
Our framework has not yet been widely applied to long-term traffic monitoring beyond this short-term experiment, and practical shortcomings remain to be explored. Additionally, effective update strategies require continually retaining historical data, posing challenges to memory limitations and data storage solutions.

For future work, we will integrate a richer set of priors to significantly improve the accuracy and reliability of seismic monitoring for vehicular activities. 
For storage solutions, we plan to develop update strategies that are more memory-efficient and can enhance monitoring performance.
Additionally, we will test our vehicle monitoring framework by deploying it over a larger network of DAS arrays and extending the analysis period, providing a more robust evaluation of its usefulness in practical scenarios. 

While our study focused on the application of DAS technology to vehicle monitoring, integrating DAS with other forms of remote sensing data could provide a more comprehensive understanding of urban dynamics and enhance intelligent city initiatives. Furthermore, accurately detecting vehicular motion along roads can improve controlled-source monitoring of seismic velocity changes beneath the DAS fiber \cite{yang2024frequency}.

\section{Conclusion}\label{sec_con}
To enhance the monitoring of seismic signals produced by urban traffic in DAS arrays, this study presents a comprehensive framework that combines a novel semi-supervised detection algorithm with advanced preprocessing methods.
In car tracing detection, we have improved the semi-supervised model Efficient Teacher by incorporating a novel shape prior loss for the same vehicle at varying speeds.
This loss term can guide the gradient direction and enhance detection accuracy without additional data.
Unlike traditional image preprocessing techniques, our approach integrates 1D vehicle detection algorithms (STA/LTA) at the preprocessing stage, reducing interfering noise. 
Additionally, our framework includes autonomous updates that better reflect real-world conditions.
Our experimental findings offer new perspectives on choosing optimal update strategies.


%



\ifCLASSOPTIONcaptionsoff
  \newpage
\fi

\bibliography{ref}
\bibliographystyle{IEEEtran}

\end{document}